\documentclass{article}
\usepackage{graphicx} % Required for inserting images
\usepackage{xcolor}
\usepackage{hyperref}
\usepackage{natbib}
\usepackage{amsmath}
\usepackage{amsfonts}
\usepackage{amsthm}
    \newtheorem{defn}{Definition}[section]
    \newtheorem{prop}{Proposition}[section]
    \newtheorem{ass}{Assumption}[section]
    \newtheorem{ex}{Example}[section]
    \newtheorem{thm}{Theorem}[section]
    \newtheorem*{coro}{Corollary}
    
\usepackage{enumitem}
    \setlist[enumerate]{itemsep=0pt, labelsep=5pt, topsep=2pt, parsep=0pt}
\usepackage[most]{tcolorbox}
\setcitestyle{aysep={}}

\title{
\vspace*{-3cm}
Substitutability, equilibrium transport,\\ and matching models%
\thanks{Support from the European Research Council Consolidator Grant EQUIPRICE (ERC-CoG No.~866274) is gratefully acknowledged.}}

\author{
Alfred Galichon\thanks{New York University, New York, and Sciences Po, Paris, \texttt{alfred.galichon@nyu.edu}}
\and
Antoine Jacquet\thanks{Sciences Po, Paris, \texttt{antoine.jacquet@sciencespo.fr}}
}

\date{April 2024}

\begin{document}

\maketitle

\begin{abstract}
\noindent This chapter surveys how \emph{substitutability} organizes computation in a large class of models encompassing optimal transport and matching.
We review Z-, M\textsubscript{0}-, and M-functions and inverse-isotonicity results, showing when Jacobi's algorithm (including Sinkhorn/IPFP) converge.
For transferable and imperfectly transferable settings, a distance-to-frontier formulation recasts equilibria as M- or M\textsubscript{0}-function fixed points, clarifying existence and uniqueness.
For non-transferable utility, we reinterpret Gale–Shapley and Adachi, expose the lattice of stable matchings, and present DALM for large markets.
Toy applications--hedonic surge pricing and rent-controlled housing--ground the methods.
Overall, substitutability provides a powerful alternative to convexity for theory and computation.
\end{abstract}

\section{Introduction}

This chapter is about \emph{substitutability}.
Substitutability is a fundamental property of optimal transport models, although it is less recognized than convexity.

Consider the problem of computing an economic equilibrium 
\begin{equation}
Q \left(p\right) = 0
\end{equation}
where $p$ is a vector of prices for $n$ goods, and $Q : \mathbb{R}^n \rightarrow \mathbb{R}^n$ is an excess supply function (supply minus demand for each of the $n$ goods).

Broadly speaking, there are only two categories of economic models that can be conveniently computed.
The first category includes the models that rely on convexity, i.e.\ models where $Q = \nabla U \left(p\right)$ with $U$ a convex function.
These models can be reformulated as an optimization problem (minimization of $U$) and \emph{descent methods}, such as the standard gradient descent, will work: 
\begin{equation}
p_z^{t+1} = p_z^t - \varepsilon \, Q_z \left(p^t\right), \quad \text{$\varepsilon > 0$ small}.
\end{equation}

The second category comprises the models that rely on substitutability -- essentially the idea that $Q_z$ is increasing in $p_z$ and decreasing in $p_x$ for $x \neq z$.
In this case, \emph{coordinate update methods}, such as nonlinear Jacobi, where we set the price of good $z$ to clear the market for good $z$, will work:
\begin{equation}
Q_z \left(p_z^{t+1}, p_{-z}^t\right) = 0.
\end{equation}

\bigskip

By an extraordinary coincidence, optimal transport inherits both structures.
For this reason, one can compute the OT problem and its regularizations both by descent methods and by coordinate update methods (which is called Sinkhorn's algorithm).

\bigskip

Some matching models can be computed using optimal transport.
These are called transferable utility models, and they assume that everyone's valuations are expressed in the same monetary unit.
In that case, equilibrium in matching problems is the solution to an optimization problem.
However, in many other cases this assumption cannot be made.
In labor economics, for instance, one dollar for the firm does not have the same value as one dollar for the employee (pre-tax or post-tax; decreasing marginal utility, etc).
Similarly, in family economics partners may transfer utility by allocating public and private expenditures that can be inefficient.
Lastly, in school choice problems utility is cardinal and cannot be transferred.

\bigskip

We will see mathematically that these models \emph{cannot} be recast as optimization problems, and so the convex optimization structure is lost.
They can, however, be reformulated as models with substitutability, from which we will be able to deduce a lot of structure and computational methods.

\bigskip

The chapter is organized as follows.
We will start in section~\ref{sec:M-functions} by introducing Jacobi's algorithm and some of its basic properties.
We will proceed with a rather detailed study of substitutability and its mathematical counterparts, \emph{Z- and M-functions}.
There was a vibrant literature on this topic in the 1970s (Birkhoff, Rheinboldt, Ortega, More, Porsching, and some others;
see in particular \citeauthor{ortega1970iterative}'s 1970 book), but this literature has somehow fallen out of attention.
We will unearth some of the main results from that literature regarding the convergence of Jacobi's algorithm, and provide some new ones.
The section ends with a toy hedonic model as an illustration.

\bigskip

In section~\ref{sec:transfer}, we will appeal to the machinery developed in section~\ref{sec:M-functions} to study models of matching with transfers, departing from the assumption of perfectly transferable utility.
The section will deal with the regularized and unregularized cases, and it will show the convergence of IPFP, and discuss the existence and uniqueness of an equilibrium.

\bigskip

In section~\ref{sec:notransfer}, we will revisit Gale and Shapley's famous theory of ``stable marriages,'' which is the theory of matchings without transfers, and we will reinterpret the well-known deferred acceptance algorithm as a damped version of Jacobi's algorithm.
We also present Adachi's formulation of the stable matching problem, which leads to the more recent notion of equilibrium matching.
%Lastly, it will examine the findings of Kelso–Crawford and Hatfield–Milgrom.

\bigskip

We hope this chapter will provide a solid mathematical introduction to matching models in economics and econometrics.

\newpage

\section{M-functions and Jacobi's algorithm}
\label{sec:M-functions}

In this section we introduce the mathematical notions related to models with substitutability, notably Z- and M-functions, as well as Jacobi's algorithm for solving such models.

\subsection{Jacobi's algorithm}

Put simply, our goal is to solve a system of nonlinear equations
\begin{equation}
Q \left(p\right) = 0 \label{eq:Qp0}
\end{equation}
where $Q : \mathbb{R}^Z \rightarrow \mathbb{R}^Z$ and $|Z| = n$.
We shall assume throughout:
\begin{ass}[Continuity]
$Q$ is continuous.
\label{ass:continuity}
\end{ass}

In economic terms, $Q$ can be interpreted as an excess supply function.
Suppose the demand for good $z$ is a function $D_z \left(p\right)$ which depends on all prices, and similarly the supply for good $z$ is a function $S_z \left(p\right)$.
Then
\begin{equation}
Q_z \left(p\right) = S_z \left(p\right) - D_z \left(p\right)
\end{equation}
is the excess supply for good $z$.
Solving $Q_z \left(p\right) = 0$ for all $z$ is thus akin to finding a vector of prices $p$ which clears the markets for all $n$ goods simultaneously.

\bigskip

Jacobi's algorithm is a method to solve such a system.
It consists of taking an initial guess $p^0$ and defining a sequence $(p^t)$ iteratively according to 
\begin{equation}
p_z^{t+1}: \quad Q_z \left( p_z^{t+1}, p_{-z}^t \right) = 0.
\label{eq:Jacobi_algorithm}
\end{equation}
Here $p_{-z}$ denotes the vector $p$ without its $z$\textsuperscript{th} entry, and $(\pi, p_{-z})$ the vector $p$ with its $z$\textsuperscript{th} entry replaced by $\pi$.
The mapping $T : p^t \mapsto p^{t+1}$ is called the \emph{coordinate update} or \emph{Jacobi update} operator.
To ensure that this operator is well-defined, we shall assume throughout:

\begin{ass}[Responsiveness]
\label{ass:responsiveness}
For all $z$,
\begin{equation}
\inf_{\pi} ~ Q_z \left(\pi, p_{-z}\right)
< 0 <
\sup_{\pi} ~ Q_z \left(\pi, p_{-z}\right),
\quad \forall p_{-z}.
\end{equation}
\end{ass}

We also remove any ambiguity in choosing $p_z^{t+1}$ by setting precisely
\begin{equation}
p_z^{t+1} = \min \big\{ \pi : Q_z(\pi, p_{-z}^t) = 0 \big\}.
\label{eq:Jacobi_algorithm_inf}
\end{equation}

Jacobi's algorithm has an intuitive economic interpretation.
Let's assume that each good $z$ has an auctioneer in charge of determining its ``right'' price $p_z$. 
At each period, the auctioneer for good $z$ sets $p_z$ in order to clear the corresponding market, but using the prices from the previous period (it is as if each auctioneer did not anticipate that other prices were being updated simultaneously).
Because of this, once the simultaneous price updates are taken into account, the price chosen by the auctioneer is in fact typically not market clearing.
This process is then re-iterated until the prices converge.
%Let's assume that each good has an auctioneer in charge of determining the ``right'' price for each good. 
%At each period, the auctioneer in charge of good $z$ sets the corresponding price $p_z$ in order to clear that market using the price information for the other goods from the previous period, not anticipating that they are simultaneously being updated, too. After the other prices have been updated, what the auctioneer for good $z$ thought was the right price is in fact typically no longer market clearing.
%This price adjustment process is then re-iterated at every period until the price adjustments fall below a certain tolerance level.

%\begin{tcolorbox}[title=Your Box Title]
%This is the content of the box.
%\end{tcolorbox}

\bigskip

As an illustration, consider the regularized optimal transport problem
\begin{align*}
\max_{\mu \geq 0} ~ &\sum_{xy \in X \times Y} \mu_{xy} \Phi_{xy} - \sum_{xy \in X \times Y} \mu_{xy} \log \mu_{xy} \\
\text{s.t.}~ &\sum_{y \in Y} \mu_{xy} = n_x, \quad \sum_{x \in X} \mu_{xy} = m_y,
\end{align*}
which has solutions of the form: $\mu_{xy} = \exp \left(\Phi_{xy} - u_x - v_y \right)$.
A common interpretation is that of a two-sided matching market between workers $x$ and firms $y$ who respectively ask for utility $u_x$ and $v_y$.
We will come back to this problem and its interpretation in section \ref{sec:transfer}.

A standard procedure to solve this type of problem is Sinkhorn's algorithm, which is nothing else than Jacobi's algorithm adapted to this special case.

\bigskip

\noindent\rule{\textwidth}{0.4pt}
\textbf{Sinkhorn's algorithm}

Take an initial guess for $u^0$ and $v^0$, and set for $t \geq 0$:
\begin{equation*}
\left\{ \;
\begin{aligned}[c]
u_x^{t+1} &: \quad n_x = \textstyle\sum_y \exp \left( \Phi_{xy} - u_x^{t+1} - v_y^t \right) \\[5pt]
v_y^{t+1} &: \quad m_y = \textstyle\sum_x \exp \left( \Phi_{xy} - u_x^t - v_y^{t+1} \right)
\end{aligned}
\right.
\end{equation*}
until approximate convergence is reached.
(Notice that here, $u^{2t}$ and $v^{2t+1}$ depend exclusively on the starting point $u^0$, while $u^{2t+1}$ and $v^{2t}$ depend exclusively on the starting point $v^0$.)

\noindent\rule{\textwidth}{0.4pt}

\bigskip

It is easy to see that by defining $p_x = -u_x$, $p_y = v_y$, as well as
\begin{align*}
Q_x \left(p\right) &= \textstyle\sum_y \exp \left(\Phi_{xy} + p_x - p_y \right) - n_x \\[0.0em]
Q_y \left(p\right) &= -\textstyle\sum_x \exp \left(\Phi_{xy} + p_x - p_y \right) + m_y,
\end{align*}
we can rewrite Sinkhorn's algorithm exactly as (\ref{eq:Jacobi_algorithm}).

Note that $Q_x$ is an antitone function with respect to $p_{-x}$ (i.e.\ nonincreasing with respect to the componentwise order), and similarly $Q_y$ is an antitone function with respect to $p_{-y}$. 
We will see below how this property plays an important role in Jacobi's algorithm.
%in the first part of the update, $u^{t+1}$ is an antitone function of $v^t$ (i.e.\ nonincreasing with respect to the componentwise order),
%while in the second part of the update, $v^{t+1}$ is an antitone function of $u^t$.

\bigskip

One important question remains: does the Jacobi sequence $(p^t)$ converge?
Although this is not the case in general, this sequence does converge when the model exhibits several features related to substitutability.
The rest of this section will be dedicated to defining exactly what we mean by substitutability, and to studying this category of models -- and notably the behavior of the Jacobi sequence.

%Several questions arise about this process.
%First, is $p^{t+1}_z$ larger than $p^t_z$?
%Second, is $p^{t+1}$ still a subsolution assuming $p^t$ is one? Third, does the sequence $\left(p^t\right)$ converge towards a solution?

%In the next paragraph, we will introduce the relevant assumptions on $Q$ in order to be able to answer these questions affirmatively.
%The core assumptions will essentially rely on the economic notion of \emph{substitutes}.

\subsection{Nondecreasing Jacobi sequence}

The convergence of the Jacobi sequence will rely on the notion of subsolution, that we define as follows:

\begin{defn}
We say that $p$ is a subsolution if $Q \left(p\right) \leq 0$,
and that $p$ is a supersolution if $Q \left(p\right) \geq 0$.

%A solution is both a sub- and a supersolution, i.e.\ $Q \left(p\right) = 0$.
\end{defn}

Broadly speaking, we expect to interpret a subsolution as ``prices are too low,'' and a supersolution as ``prices are too high.''
We shall develop this intuition below.

\bigskip

Let $p^t$ be a subsolution and assume $p^{t+1}$ exists, so that
\begin{equation}
Q_z \left(p_z^t, p_{-z}^t\right) \leq 0,
\qquad
Q_z \left(p_z^{t+1}, p_{-z}^t\right) = 0.
\end{equation}
Then we can guarantee that $p^t \leq p^{t+1}$ using:

\begin{ass}[Diagonal isotonicity]
$Q_z \left(p\right)$ is nondecreasing in $p_z$.
\label{ass:diagonal_isotonicity}
\end{ass}

%This assumption allows us to redefine the Jacobi coordinate updating as
%\begin{equation}
%p_z^t = \inf \left\{ \pi : Q_z \left( \pi, p_{-z}^t\right) \geq 0\right\}
%\end{equation}%
%where the infimum is taken in $\mathbb{R} \cup \{ \pm\infty \}$ (with the convention $\inf \emptyset = +\infty$).

We state this result formally:

\begin{prop}
\label{prop:nondecreasing_Jacobi_update}
Assume $Q$ is diagonal isotone.
If $p^t$ is a subsolution, then $p^t \leq p^{t+1}$.
\end{prop}

\begin{proof}
$Q_z \left(p_z^t, p_{-z}^t\right) \leq 0 = Q_z \left(p_z^{t+1}, p_{-z}^t\right)$ and $Q_z$ is increasing in $p_z$.
\end{proof}

\bigskip

%Diagonal isotonicity thus ensures that $p^t \leq p^{t+1}$ as long as $p^t$ is a subsolution.
Next, we can guarantee that $p^{t+1}$ is still a subsolution with: 

\begin{ass}[Z-function]
$Q_z \left(p_z, p_{-z}\right)$ is antitone with respect to $p_{-z}$,
i.e.\ $p_{-z} \leq p_{-z}'$ implies $Q_z \left(p_z, p_{-z}\right) \geq Q_z \left(p_z, p_{-z}'\right)$.
\label{ass:Z-function}
\end{ass}

In economic terms, $Q$ being a Z-function means that when the price of any other good increases, the excess supply for good $z$ should decrease.
This can be explained by a standard substitution pattern: if all prices but $p_z$ increase, then producers should supply less good $z$, while consumers should demand more good $z$.
For this reason, we also say that $Q$ has the \emph{substitutes property}.

With the Z-function assumption we have:

\begin{prop}
\label{prop:subsolution_stability}
Assume $Q$ is a diagonal isotone Z-function.
If $p^t$ is a subsolution, then $p^{t+1}$ is also a subsolution.
\end{prop}

\begin{proof}
We have $p^t \leq p^{t+1}$ by diagonal isotonicity, so in particular $p_{-z}^t \leq p_{-z}^{t+1}$ for any $z$.
Hence, since $Q_z$ is antitone with respect to $p_{-z}$, $Q_z \left( p_z^{t+1}, p_{-z}^{t+1} \right) \leq Q_z \left( p_z^{t+1}, p_{-z}^t \right) = 0$.
\end{proof}

The immediate consequence of propositions \ref{prop:nondecreasing_Jacobi_update} and \ref{prop:subsolution_stability} is that, with $Q$ satisfying the assumptions above, any Jacobi sequence starting from a subsolution will be nondecreasing.
Yet, as the following example shows, this is still not sufficient to ensure convergence.

\begin{ex}
\label{ex:Jacobi_diverges}
Consider a linear $Q \left(p\right)$ given by
\begin{equation*}
Q \left(p\right) = \begin{pmatrix}
1 & -2 \\ 
-2 & 1
\end{pmatrix}
p.
\end{equation*}
The Jacobi sequence starting from $p^0$ can be written explicitly as
\begin{equation*}
\left\{ \;
\begin{aligned}[c]
p_1^t &= 2^t \, p_2^0 \\
p_2^t &= 2^t \, p_1^0.
\end{aligned}
\right.
\end{equation*}
Even though $p^0 = (1,1)$ is a subsolution, and $Q$ is a continuous diagonal isotone Z-function, the associated Jacobi sequence does not converge to the unique solution $(0,0)$.
In fact, it is clear from the expressions of $p_1^t$ and $p_2^t$ that any Jacobi sequence starting from a vector other than $(0,0)$ will diverge.
\end{ex}

\subsection{M- and M\textsubscript{0}-functions}

Example \ref{ex:Jacobi_diverges} shows what went wrong:
the function $Q$ allowed for an inversion by allowing the subsolution $(1,1)$ to be above the solution $(0,0)$.
Recall that we expect subsolution to mean ``prices are too low'' and supersolution ``prices are too high,'' and therefore we wish to rule out such inversions.
We do so using the following notions, defined in \citet*{galichon2022monotone}:

\begin{defn}
Consider $Q : \mathbb{R}^n \rightarrow \mathbb{R}^n$.
We say that: \begin{enumerate}

\item[(i)] $Q$ is strongly nonreversing if 
\begin{equation}
\left\{ 
\begin{array}{l}
p \geq p' \\ 
Q \left(p\right) \leq Q \left(p'\right)
\end{array}
\right. \text{ implies }
p = p'.
\end{equation}

\item[(ii)] $Q$ is weakly nonreversing if 
\begin{equation}
\left\{ 
\begin{array}{l}
p \geq p' \\
Q \left(p\right) \leq Q \left(p'\right)
\end{array}
\right. \text{ implies }
Q \left(p\right) = Q \left(p'\right).
\end{equation}

\end{enumerate}

\end{defn}

\paragraph{}
Economically speaking, $Q$ being nonreversing means that when prices increase, the excess supply cannot decrease for all goods.
In other words, it ensures that the price effect remains stronger than the substitution effect.

\bigskip

\begin{ex}
Consider $Q \left(p\right) = Qp$ with $Q$ an $n \times n$ matrix.
If $Q$ is weakly (column) diagonally dominant (i.e.\ $1^\top Q \geq 0$), then $Q$ is weakly nonreversing.
If it is strictly (column) diagonally dominant (i.e.\ if $1^\top Q > 0$), then $Q$ is strongly nonreversing.
\end{ex}

\begin{ex}
Consider $Q$ such that $1^\top Q \left(p\right)$ is weakly isotone in $p$.
Then it is weakly nonreversing.
In particular, this is the case of $Q : \mathbb{R}^{X\cup Y} \rightarrow \mathbb{R}^{X\cup Y}$ given by
\begin{equation}
\left\{ \;
\begin{aligned}
Q_x \left(p\right) &= \textstyle\sum_y \exp \left( \Phi_{xy} + p_x - p_y \right) - n_x,
\quad x \in X, \\
Q_y \left(p\right)
&= -\textstyle\sum_x \exp \left( \Phi_{xy} + p_x - p_y \right) + m_y,
\quad y \in Y.
\end{aligned}
\right.
\end{equation}

\end{ex}

\begin{ex}
Consider $Q$ such that $1^\top Q \left(p\right) \leq 1^\top Q \left(p^\prime\right)$ and $p \geq p^\prime$ together imply $p = p^\prime$.
Then it is strongly nonreversing.
In particular, this is the case of $Q : \mathbb{R}^{X\cup Y} \rightarrow \mathbb{R}^{X\cup Y}$ given by
\begin{equation*}
\left\{ \;
\begin{aligned}
Q_x \left(p\right)
&= \textstyle\sum_y \exp \left( \Phi_{xy} + p_x - p_y \right) + \exp \left(p_x\right) - n_x,
&&x \in X \\
Q_y \left(p\right)
&= -\textstyle\sum_x \exp \left( \Phi_{xy} + p_x - p_y \right) - \exp \left(-p_y\right) + m_y,
&&y \in Y.
\end{aligned}
\right.
\end{equation*}
\end{ex}

\bigskip

Next we define M- and M\textsubscript{0}-functions as Z-functions which are also nonreversing:

\begin{defn}
An M-function is a Z-function which is strongly nonreversing.
An M\textsubscript{0}-function is a Z-function which is weakly nonreversing.
\end{defn}

Clearly, M-functions constitute a subset of M\textsubscript{0}-functions since strong nonreversingness implies weak nonreversingness.
In economic terms, M- and M\textsubscript{0}-functions exhibit both the substitutes property (which they inherit from Z-functions), and the property of having a price effect stronger than the substitution effect.

The combination of these two properties implies that M\textsubscript{0}-functions (and therefore M-functions also) are diagonal isotone, so that M- and M\textsubscript{0}-functions satisfy assumptions \ref{ass:diagonal_isotonicity} and \ref{ass:Z-function}.

\begin{prop}
Assume $Q$ is an M\textsubscript{0}-function, then it is diagonal isotone.
\end{prop}

\begin{proof}
Let $p$ and $p'$ such that $p_z \leq p_z'$ and $p_{-z} = p_{-z}'$.
Looking for a contradiction, suppose that $Q_z \left(p\right) > Q_z \left(p'\right)$.

For any $\tilde z \neq z$ we have $p_{\tilde z} = p_{\tilde z}'$ and $p_{-\tilde z} \leq p_{-\tilde z}'$, so since $Q$ is a Z-function,
\begin{equation*}
Q_{\tilde z} \left(p\right) = Q_{\tilde z} \left( p_{\tilde z}, p_{-\tilde z} \right) \geq Q_{\tilde z} \left( p_{\tilde z}, p_{-\tilde z}' \right) = Q_{\tilde z} \left(p'\right).
\end{equation*}
Hence $p \leq p'$ and $Q \left(p\right) \geq Q \left(p'\right)$, therefore $Q \left(p\right) = Q \left(p'\right)$ by weak nonreversingness.
We have a contradiction: in fact $Q_z \left(p\right) \leq Q_z \left(p'\right)$.
\end{proof}

\bigskip

\citet*{galichon2022monotone} characterize M-functions using:

\begin{thm}[Inverse isotonicity theorems]
\label{thm:inverse_isotonicity}
Consider $Q$ a Z-function. Then: \begin{enumerate}
\item[(i)] $Q$ is an M-function if and only if it is inverse isotone, i.e.
\begin{equation*}
Q \left(p\right) \leq Q \left( p^\prime \right) \text{ implies } p \leq p^\prime.
\end{equation*}
\item[(ii)] $Q$ is an M\textsubscript{0}-function if and only if the set-valued map $Q^{-1}$ is isotone in the strong set order, i.e.\ 
\begin{equation*}
Q \left(p\right) \leq Q \left( p' \right) \text{ implies } \left\{
\begin{aligned}
&Q \left(p\right) = Q \left( p \wedge p' \right) \\
&Q \left(p'\right) = Q \left( p \vee p' \right).
\end{aligned}
\right.
\end{equation*}
\end{enumerate}
\end{thm}

\bigskip

We recall that $p \vee p'$ denotes the \emph{join} (componentwise maximum) of two vectors $p$ and $p'$, while $p \wedge p'$ denotes their \emph{meet} (componentwise minimum).

\begin{proof}
(i) Assume $Q$ is an M-function and $Q \left(p\right) \leq Q \left( p' \right)$.
Then $p \vee p' \geq p'$ and
\begin{equation*}
Q_z \left( p \vee p' \right)
\; \leq \; 1\left\{ p_z \leq p_z' \right\} Q_z \left( p' \right)
+ 1 \left\{ p_z > p_z' \right\} Q_z \left(p\right)
\; \leq \; Q_z \left( p' \right),
\end{equation*}
therefore we conclude that $p' = p \vee p'$, and thus $p \leq p'$.
Conversely, assume $Q$ is inverse isotone.
To show it is strongly nonreversing, assume $p \geq p'$ and $Q \left(p\right) \leq Q \left( p' \right)$.
By inverse isotonicity $p \leq p'$, hence $p = p'$.

(ii) Assume $Q$ is an M\textsubscript{0}-function and $Q \left(p\right) \leq Q \left( p' \right)$.
Then $p \vee p' \geq p'$ and
\begin{equation*}
Q_z \left( p \vee p' \right)
\leq 1 \left\{ p_z \leq p_z' \right\} Q_z \left( p' \right)
+ 1 \left\{ p_z > p_z' \right\} Q_z \left(p\right)
\leq Q_z \left( p' \right),
\end{equation*}
therefore we conclude that $Q \left( p \vee p' \right) = Q \left( p' \right)$.
Similarly, $p \geq p \wedge p'$ and
\begin{equation*}
Q_z \left( p \wedge p' \right)
\geq 1 \left\{ p_z \leq p_z^\prime \right\} Q_z \left(p\right)
+ 1 \left\{ p_z > p_z' \right\} Q_z \left( p' \right)
\geq \; Q_z \left(p\right) 
\end{equation*}
thus $Q \left( p \wedge p' \right) = Q \left(p\right)$.
Conversely, assume $Q^{-1}$ is isotone in the strong set order.
To show that $Q$ is weakly nonreversing, assume $p \geq p'$ and $Q \left(p\right) \leq Q \left( p' \right)$.
The former implies $p \wedge p' = p'$, and the latter $Q \left(p\right) = Q \left( p \wedge p' \right)$.
\end{proof}

\begin{coro}
Assume $Q$ is an M\textsubscript{0}-function,
then the set of solutions is stable by the join and meet operations $\vee$ and $\wedge$.
%Then the set of subsolutions is stable by the meet operation $\vee$,
%and the set of supersolutions is stable by the join operation $\wedge$.
%Hence the set of solutions is stable by both operations.
%and that both a sub- and a supersolution exist.
%Then the set of solutions is stable by the join and meet operations $\wedge$ and $\vee$.
\end{coro}

%\begin{proof}
%$Q \left(p\right) = Q \left( p' \right) = 0$ implies $Q \left( p \vee p' \right) \leq 0 \leq Q \left( p \wedge p' \right)$,
%but since $p \vee p' \geq p \wedge p'$ and $Q$ is weakly nonreversing, $Q \left( p \wedge p' \right) = Q \left( p \vee p^\prime \right) = 0$.
%\end{proof}

\subsection{Convergence of the Jacobi sequence}

When $Q$ is an M- or M\textsubscript{0}-function, inversions such as the one we saw in example \ref{ex:Jacobi_diverges} are ruled out.
In this case, as long as a sub- and a supersolution exist, we can find a Jacobi sequence which converges towards a solution.

\bigskip

\cite{ortega1970iterative} prove the two following results for M-functions.

\begin{thm}
\label{thm:Ortega-Rheinboldt1970}
Assume $Q$ is a continuous responsive M-function, and that both a sub- and a supersolution exist.
Then a solution exists and it is unique, and any Jacobi sequence starting from a sub- or supersolution converges to this unique solution.
\end{thm}

\begin{proof}
The proof consists in considering two Jacobi sequences $(\hat p^t)$ and $(\check p^t)$ starting from respectively a sub- and a supersolution.
$(\hat p^t)$ is an increasing sequence of subsolutions, $(\check p^t)$ is a decreasing sequence of supersolutions, and by inverse isotonicity, one has $\hat p^t \leq \check p^t$, and therefore they both converge to respectively $p$ and $p'$.
$Q \left(p\right) = Q \left( p' \right) = 0$ and inverse isotonicity applied twice yields $p = p'$.
\end{proof}

\bigskip

If $Q$ is surjective, then we need not even start from a sub- or supersolution.

\begin{thm}
\label{thm:Ortega-Rheinboldt1970surjective}
Assume $Q$ is also surjective.
Then any Jacobi sequence converges to the unique solution.
\end{thm}

\begin{proof}
Let $(p^t)$ be the Jacobi sequence starting from a vector $p^0$.
Since $Q$ is surjective, there exist $\hat p^0$ such that $Q \left( \hat p^0 \right) = Q \left( p^0 \right) \wedge 0$,
and $\check p^0$ such that $Q \left( \check p^0 \right) = Q \left( p^0 \right) \vee 0$.
Then $\hat p^0$ is a subsolution, $\check p^0$ is a supersolution, and the Jacobi sequences $(\hat p^t)$ and $(\check p^t)$ starting from these respective vectors are such that 
\begin{equation*}
\hat p^t \leq p^t \leq \check p^t
\end{equation*}
with $(\hat p^t)$ increasing and $(\check p^t)$ decreasing.
As a result, all three sequences converge, and their limits are solutions.
\end{proof}

\bigskip

Finally, \cite{GalichonLegerZmappings} obtain the following result for M\textsubscript{0}-functions.
%Recall that a sublattice of $\mathbb R^n$ is a subset $L \subset \mathbb R^n$ such that for any two vectors $p, p' \in L$, the join and the meet of $p$ and $p'$ are also in $L$.

\begin{thm}
\label{thm:Galichon-Leger2023}
Assume $Q$ is an M\textsubscript{0}-function, and that both a sub- and a supersolution exist.
Then a solution exists and can be obtained as the limit of the Jacobi sequence starting from the meet (or the join) of a subsolution and a supersolution.
\end{thm}

\begin{proof}
Let $\check p$ be a subsolution and $\hat p$ a supersolution.
Then $Q \left( \check p \right) \leq Q \left( \hat p \right)$, therefore the inverse isotonicity characterization of M\textsubscript{0}-functions implies $Q \left( \check p \wedge \hat p \right) = Q \left( \check p \right)$ and $Q \left( \check p \vee \hat p \right) = Q \left( \hat p\right)$.
Hence $\check p \wedge \hat p$ is a subsolution and $\check p \vee \hat p$ is a supersolution.
Consider $(\underline p^t)$ and $(\bar p^t)$ the Jacobi sequences starting respectively from $\underline p^0 = \check p \wedge \hat p$ and $\bar p^0 = \check p \vee \hat p$, so that $\underline p^0 \leq \bar p^0$.
We show by induction that $\underline p^t \leq \bar p^t$.
Looking for a contradiction, assume that $\underline p{}_z^{t+1} > \bar p_z^{t+1}$ for some $z$.
Then
\begin{equation*}
0 = Q_z \big( \underline p{}_z^{t+1}, \underline p{}_{-z}^t \big)
\geq Q_z \big( \bar p_z^{t+1}, \underline p{}_{-z}^t \big)
\geq Q_z \big( \bar p_z^{t+1}, \bar p_{-z}^t \big) = 0
\end{equation*}
by diagonal isotonicity and the Z-function property.
Hence $Q_z \big( \underline p{}_z^{t+1}, \underline p{}_{-z}^t \big) = Q_z \big( \bar p_z^{t+1}, \underline p{}_{-z}^t \big) = 0$ but $\underline p{}_z^{t+1} > \bar p_z^{t+1}$, which contradicts the definition \eqref{eq:Jacobi_algorithm_inf} of $\underline p{}_z^{t+1}$ as the smallest solution to $Q_z \big( \pi, \underline p{}_{-z}^t \big) = 0$.
Thus $(\underline p^t)$ is increasing, $(\bar p^t)$ is decreasing, and $\underline p^t \leq \bar p^t$, therefore both sequences are bounded and have a limit.
Hence a solution exists.
\end{proof}

\bigskip

We now consider some examples around \emph{constant aggregates},
i.e.\ the property that $1^\top Q \left(p\right) = \sum_z Q_z \left(p\right)$ is constant for all $p$.

\begin{prop}
Assume $Q$ is a continuous Z-function with $1^\top Q \left(p\right) = 0$, and that there is a $0 \in Z$ such that the restriction and corestriction of $Q$ to $\mathbb{R}^{Z \backslash \{0\}}$ with the normalization $p_0 = \pi$, denoted $Q_{-0} : \mathbb{R}^{Z \backslash \{0\}} \to \mathbb{R}^{Z \backslash \{0\}}$, is a surjective M-function.
Then, denoting $p \left(\pi\right)$ the unique solution to $Q_{-0} \left(p\right) = 0$, one has that $\pi \leq \pi'$ implies $p \left( \pi \right) \leq p \left( \pi' \right)$.
\end{prop}

\bigskip

\begin{ex}
Consider the case of regularized optimal transport:
\begin{equation*}
\left\{ \;
\begin{aligned}[c]
Q_x \left(p\right) &= \textstyle\sum_y \exp \left( \Phi_{xy} + p_x - p_y \right) - n_x \\
Q_y \left(p\right) &= -\textstyle\sum_x \exp \left( \Phi_{xy} + p_x - p_y \right) + m_y,
\end{aligned}
\right.
\end{equation*}
so that $1^\top Q \left(p\right) = \sum_y m_y - \sum_x n_x$ is constant.
It is easy to see that $Q$ is a Z-function.
In addition, because of constant aggregates, $Q \left(p\right) \leq Q \left(p'\right)$ implies $Q \left(p\right) = Q \left(p'\right)$ so $Q$ is weakly nonreversing, thus it is an M\textsubscript{0}-function.
However, $Q$ is not an M-function: take for instance $p = p' + 1_Z$, then we have $p \geq p'$ and $Q \left(p\right) = Q \left(p'\right)$, but $p \neq p'$ so $Q$ is not strongly nonreversing.
\end{ex}

\bigskip

\begin{ex}
Consider a linear $Q \left(p\right) = (\Delta - A) p$ where $\Delta$ is a diagonal matrix with positive entries,
$A$ has a zero diagonal and positive off-diagonal entries, and
\begin{equation*}
1^\top (\Delta - A) = 0
\end{equation*}
so that the function $Q$ has constant aggregates.
Then the matrix $\Delta^{-1} A$ is nonnegative and irreducible (it is strongly connected), and the Perron–Frobenius theorem applies.
Letting $\delta = \Delta 1$, we have $\delta \geq 0$ and $\delta^\top = 1^\top A = \delta^\top \Delta^{-1} A$, so $\delta$ must be the left Perron eigenvector of $\Delta^{-1} A$, and it is associated to the Perron eigenvalue 1.
Thus there is also a right Perron eigenvector $v \geq 0$ such that $\Delta^{-1} A v = v$, i.e.\ $Q \left(v\right) = (\Delta - A) v = 0$.

%Constant aggregates is not enough to ensure that $\Delta - A$ has connected strong substitutes, i.e.\ that $\Delta^{-1} A$ is irreducible.
%Consider for instance
%\begin{equation*}
%\Delta - A =
%\begin{pmatrix}
%1 & -1 & 0 & 0 \\
%-1 & 1 & 0 & 0 \\
%0 & 0 & 1 & -1 \\
%0 & 0 & -1 & 1
%\end{pmatrix}.
%\end{equation*}
%Then $1^\top (\Delta - A) = 0$, but $\Delta - A$ does not have connected strong substitutes,
%so $\Delta^{-1} A$ is not irreducible.
\end{ex}

\subsection{Application: a toy hedonic model}

Consider a surge pricing problem in an Uber-like environment.
We have partitioned the city in a finite number of locations (say, blocks).
Let $x \in X$ denote the location of the driver, $y \in Y$ the location of the passenger, and $z \in Z$ the pickup location.
Assume that for a driver at $x$, the cost of picking up a passenger at $z$ is $c_{xz}$.
If the price of the ride at $z$ is $p_z$, the utility of the driver is $p_z - c_{xz} + \varepsilon_z$, where the vector $(\varepsilon_z)$ is random.
If the driver does not pick up anyone, the utility is normalized to $\varepsilon_0$.
Assume that $(\varepsilon_z)$ is i.i.d.\ Gumbel.
Then the probability that a driver at $x$ will demand a ride $z$ is 
\begin{equation}
\frac{\exp \left( p_z - c_{xz} \right)}{1 + \sum_{z'} \exp \left( p_{z'} - c_{xz'} \right)}.
\end{equation}

Now assume there are $n_x$ drivers in area $x$, therefore the supply for rides at $z$ is 
\begin{equation}
S_z \left(p\right) = \sum_{x \in X} n_x \frac{\exp \left( p_z - c_{xz} \right)}{1 + \sum_{z'} \exp \left( p_{z'} - c_{xz'} \right)}.
\end{equation}
It is easy to see that $S_z \left(p\right)$ is decreasing with respect to $p_{z'}$ for $z' \neq z$.

\bigskip

Now let's focus on demand.
This is the same as before, except for the fact that the utility of a passenger at $y$ seeking a ride at location $z$ is now $a_{yz} - p_z + \eta_z$, and the utility of the outside option is $\eta_0$, where $(\eta_z)$ is i.i.d.\ Gumbel.
Assuming there are $m_y$ passengers in area $y$, the induced demand is 
\begin{equation}
D_z \left(p\right) = \sum_{y \in Y} m_y \frac{\exp \left( a_{yz} - p_z \right)}{1 + \sum_{z'} \exp \left( a_{yz'} - p_{z'} \right)}
\end{equation}
and we see that $-D_z \left(p\right)$ is also decreasing with respect to $p_{z'}$ for $z' \neq z$.

\bigskip

The excess supply function $Q$ is defined with
\begin{equation*}
Q_z \left(p\right) = S_z \left(p\right) - D_z \left(p\right).
\end{equation*}
It is clearly continuous, and it is a Z-function as the sum of two Z-functions.
Let's show that it is also strongly nonreversing, and therefore an M-function.
The aggregates of $Q \left(p\right)$ are
\begin{equation*}
\sum_z Q_z \left(p\right) =
\sum_{x \in X} n_x \frac{\sum_z \exp(p_z - c_{xz})}{1 + \sum_z \exp(p_z - c_{xz})}
- \sum_{y \in Y} m_y \frac{\sum_z \exp(a_{yz} - p_z)}{1 + \sum_z \exp(a_{yz} - p_z)},
\end{equation*}
which we see are strictly isotone in $p$: $p \geq p'$ implies $\sum_z Q_z \left(p\right) \geq \sum_z Q_z \left(p'\right)$, and if $p \neq p'$ then the inequality is strict.
Suppose then that we have $p \geq p'$ and $Q \left(p\right) \leq Q \left(p'\right)$.
Because aggregates are isotone, we must have $Q \left(p\right) = Q \left(p'\right)$.
Furthermore, if we had $p \neq p'$, then we would have $\sum_z Q_z \left(p\right) > \sum_z Q_z \left(p'\right)$.
But this contradicts $Q \left(p\right) = Q \left(p'\right)$, hence actually $p = p'$.

\bigskip

Finally, note that when all $p_z = k$ and $k \rightarrow +\infty$, $Q_z \left(p\right) \geq 0$;
while when all $p_z = k$ and $k \rightarrow -\infty$ we have $Q_z \left(p\right) \leq 0$.
As a result both a super- and a subsolution exist, and therefore theorem \ref{thm:Ortega-Rheinboldt1970} guarantees that the problem has a unique solution which can be reached as the limit of a Jacobi sequence starting from any sub- or supersolution.

\subsection{References and notes}

This section draws upon the book by \cite{ortega1970iterative}, the papers by \citet*{berry2013connected} and \citet*{galichon2019costly}.
%Additional references include \citet*{galichon2022monotone}.
It also builds on the working papers by \citet*{galichon2022monotone}, \citet{GalichonLegerZmappings}, and \citet*{ChooGalichonLiangWeber}.

\newpage
%\section{Lecture 2: models of matching with transfers}
\section{Models of matching with transfers}
\label{sec:transfer}

In this section we shall look specifically at models which formulate as 
\begin{equation*}
\left\{ 
\begin{array}{c}
\sum_{y \in Y} M_{xy} \left( a_x, b_y \right) + M_{x0} \left( a_x \right) = n_x \\ 
\sum_{x \in X} M_{xy} \left( a_x, b_y \right) + M_{0y} \left( b_y \right) = m_y
\end{array}
\right.
\end{equation*}
where $M_{xy} \left( a_x, b_y \right)$ is continuous and increasing in $a_x$ and $b_y$, and will stand for the number of matches between types $x$ and $y$;
and $M_{x0} \left( a_x \right)$ and $M_{0y} \left( b_y \right)$ are also continuous and will stand for the number of unmatched agents of type $x$ and $y$, respectively.

At the end of this section we will show how this framework extends to the full assignment case, i.e.\ when $M_{x0} \left( a_x \right) = M_{0y} \left( b_y \right) = 0$.
Optimal transport falls into this category, as seen with Sinkhorn's algorithm in section \ref{sec:M-functions}.

\subsection{Microfoundation of the matching model}

We will consider as illustration a matching model between workers (CEOs) and firms with taxes.
There are $n_x$ workers of type $x \in X$ and $m_y$ firms of type $y \in Y$.
Assume that firms pay their worker a gross wage $w$, from which the worker receives the net wage $N \left(w\right)$.
In typical progressive taxation fashion, $N$ is increasing, concave, and piecewise affine.
We can thus represent it as 
\begin{equation}
\label{eq:net_wage_function}
N \left(w\right) = \min_{k=0,\dots,K} \left(1-\tau_k\right) \left(w - w_k\right)
\end{equation}
where $\tau_k$ is the marginal tax rate in the $k$\textsuperscript{th} tax bracket: $\tau_0 = 0 < \tau_1 < \dots < \tau_K$.
If there are no taxes, then we simply have $N \left(w\right) = w$.

If matched with a firm $y$, a worker $x$ gets utility
\begin{equation*}
\alpha_{xy} + N (w_{xy}) + \sigma \varepsilon_y
\end{equation*}
where $\alpha_{xy}$ is worker $x$'s monetary valuation of job $y$'s amenities;
$N(w_{xy})$ is the net wage of worker $x$ working for firm $y$;
and $\varepsilon_y$ is a random utility.
If unmatched, a worker $x$ gets utility $\sigma \varepsilon_0$.
Hence worker $x$'s problem is to choose a firm $y$ with
\begin{equation*}
\max_{y \in Y} \left\{ \alpha_{xy} + N (w_{xy}) + \sigma \varepsilon_y, \, \sigma \varepsilon_0 \right\}.
\end{equation*}
We assume that $(\varepsilon_y)$ is i.i.d.\ Gumbel.
As a result, the expected indirect utility of a worker of type $x$ is
\begin{equation*}
u_x
= \sigma \log \left( 1 + \sum_y \exp \left( \frac{\alpha_{xy} + N(w_{xy})}{\sigma} \right) \right)
\end{equation*}
and the probability that worker $x$ picks firm $y$ is
\begin{equation}
\frac{\mu_{xy}}{n_x}
= \frac{\exp \big( \frac{\alpha_{xy} + N (w_{xy})}{\sigma}\big)}{1 + \sum_{y'} \exp \big( \frac{\alpha_{xy'} + N (w_{xy'}) }{\sigma} \big)} 
= \exp \left( \frac{\alpha_{xy} + N (w_{xy}) - u_x}{\sigma} \right).
\label{eq:prob_x_picks_y}
\end{equation}

\bigskip

On firm $y$'s side, the problem is to choose a worker $x$ with
\begin{equation*}
\max_{x \in X} \left\{ \gamma_{xy} - w_{xy} + \sigma \eta_x, \, \sigma \eta_0 \right\}
\end{equation*}
where $\gamma_{xy}$ is the value of worker $x$ for firm $y$, and $\eta_x$ is the random utility.
We assume that $(\eta_x)$ is also i.i.d.\ Gumbel, so that the expected indirect utility of a firm of type $y$ is 
\begin{equation*}
v_y = \sigma \log \left( 1 + \sum_x \exp \left( \frac{\gamma_{xy} - w_{xy}}{\sigma}\right) \right)
\end{equation*}
and the probability that firm $y$ picks worker $x$ is
\begin{equation}
\frac{\mu_{xy}}{m_y} = \exp \left( \frac{\gamma_{xy} - w_{xy} - v_y}{\sigma} \right).
\label{eq:prob_y_picks_x}
\end{equation}
Lastly, the probabilities that worker $x$ or firm $y$ remains unmatched are respectively:
\begin{equation}
\frac{\mu_{x0}}{n_x} = \exp \left( -\frac{u_x}{\sigma} \right),
\qquad
\frac{\mu_{0y}}{m_y} = \exp \left( -\frac{v_y}{\sigma} \right).
\label{eq:prob_unmatched}
\end{equation}

\bigskip

Equations (\ref{eq:prob_x_picks_y}), (\ref{eq:prob_y_picks_x}) and (\ref{eq:prob_unmatched}), together with the feasibility conditions
\begin{equation}
\sum_{y \in Y} \mu_{xy} + \mu_{x0} = n_x,
\qquad
\sum_{x \in X} \mu_{xy} + \mu_{0y} = m_y,
\label{eq:feasibility_x_and_y}
\end{equation}
form a system of equations with unknowns $\mu_{xy}$, $\mu_{x0}$, $\mu_{0y}$, $u_x$, $v_y$ and $w_{xy}$.
%To summarize, the equations of the model are for now:
%\begin{align}
%&\textstyle\sum_{y \in Y} \mu_{xy} + \mu_{x0} = n_x, 
%&&\textstyle\sum_{x \in X} \mu_{xy} + \mu_{0y} = m_y \\
%&n_x \exp \left( \tfrac{\alpha_{xy} + N \left(w_{xy}\right) - u_x}{\sigma} \right) = \mu_{xy}
%&&m_y \exp \left( \frac{\gamma_{xy} - w_{xy} - v_y}{\sigma}\right) = \mu_{xy} \\
%&n_x \exp \left( -\tfrac{u_x}{\sigma} \right) = \mu_{x0}
%&&m_y \exp \left( -\tfrac{v_y}{\sigma} \right) = \mu_{0y}.
%\end{align}
Our strategy to solve this system will consist of three steps.
First, we will eliminate $w_{xy}$ in order to express $\mu_{xy}$ as a function of $u_x$ and $v_y$ only, using the \emph{distance-to-frontier function}.
Second, we will solve for $u_x$ and $v_y$ with Jacobi's algorithm.
Finally, we will recover $\mu_{xy}$, $\mu_{x0}$, $\mu_{0y}$ and $w_{xy}$ from $u_x$ and $v_y$.

\subsection{Distance-to-frontier function}

The wage $w_{xy}$ is a way to transfer systematic utility from one partner to the other.
After transfer, the systematic utilities of both partners in a match $xy$ are
\begin{equation*}
\left\{ \;
\begin{aligned}
&\text{for $x$,} &&U_{xy} = \alpha_{xy} + N \left(w_{xy}\right) \\
&\text{for $y$,} &&V_{xy} = \gamma_{xy} - w_{xy}.
\end{aligned}
\right. 
\end{equation*}

\bigskip

More generally, we will consider models in which $(U,V)$ belongs to a feasible set $\mathcal{F}_{xy}$, and we shall make two assumptions on $\mathcal{F}_{xy}$.
The first one, \emph{free disposal}, means that agents can dispose of utility at will.

\begin{ass}[Free disposal]
If $(U,V) \in \mathcal{F}_{xy}$, then for any $U' \leq U$ and $V' \leq V$ we have $(U',V') \in \mathcal{F}_{xy}$.
\end{ass}

The second one, \emph{scarcity}, means that one partner cannot have an arbitrarily large level of utility without a negative effect on the utility of the other partner.

\begin{ass}[Scarcity]
If $(U^n)$ and $(V^n)$ are two sequences such that $U^n \to +\infty$ and $V^n$ is bounded below, then for $N$ large enough $(U^n, V^n) \notin \mathcal F_{xy}$ for $n \geq N$; and similarly for $U^n$ bounded below and $V^n \to +\infty$.
\end{ass}

\bigskip

We will describe the feasible set $\mathcal{F}_{xy}$ using the \emph{distance-to-frontier function}:
\begin{equation}
D_{xy} \left(U,V\right) = \min \left\{ t \in \mathbb{R} : (U-t,V-t) \in \mathcal{F}_{xy} \right\}.
\end{equation}
This function encodes the feasible set in the following sense.
If the utility profile $(U,V)$ is outside the feasible set, then the distance is positive: $D_{xy} \left(U,V\right) > 0$.
Conversely, if $(U,V)$ is in the interior of the feasible set, then the distance is negative: $D_{xy} \left(U,V\right) < 0$.
Only when $(U,V)$ is exactly on the frontier is the distance zero.

\bigskip

We mention two other useful properties of the distance-to-frontier function, before looking at this function on two examples.

\begin{prop} ~
    \begin{enumerate}
        \item[(i)] For $F_1$ and $F_2$ two elementary sets, 
\begin{equation*}
D_{F_1 \cap F_2} = \max \left\{ D_{F_1}, D_{F_2} \right\}
\quad \text{and} \quad
D_{F_1 \cup F_2} = \min \left\{ D_{F_1}, D_{F_2} \right\}.
\end{equation*}
        \item[(ii)] Translation invariance: $D \left(U + t, V + t\right) = t + D \left(U,V\right)$.
    \end{enumerate}
\end{prop}

\bigskip

\begin{ex}[Transferable utility]
In this case
\begin{equation*}
\mathcal{F}_{xy} = \left\{ (U,V) \in \mathbb{R}^2 : U + V \leq \Phi_{xy} \right\},
\end{equation*}
so the minimum $t$ such that $(U-t,V-t) \in \mathcal{F}_{xy}$ verifies $(U-t) + (V-t) = \Phi_{xy}$,
hence
\begin{equation*}
D_{xy} \left(U,V\right) =\frac{U + V - \Phi_{xy}}{2}.
\end{equation*}
\end{ex}

\begin{ex}[Piecewise linear taxes]
In this case $U \leq \alpha_{xy} + N \left(w_{xy}\right)$ and $V \leq \gamma_{xy} - w_{xy}$, so that
\begin{align*}
\mathcal{F}_{xy}
&= \left\{ (U,V) \in \mathbb{R}^2 : N \left( \gamma_{xy} - V \right) \geq U - \alpha_{xy} \right\} \\
&= \big\{ (U,V) : \min_{k=0,\dots,K} (1-\tau_k) (\gamma_{xy} - V - w_k) \geq U - \alpha_{xy} \big\} \\
&= \bigcap_k \big\{ (U,V) : (1-\tau_k) (\gamma_{xy} - V - w_k) \geq U -\alpha_{xy} \big\}.
\end{align*}
Hence $D_{xy} \left(U,V\right) = \max_k ~ D_{xy}^k \left(U,V\right)$, where 
\begin{equation*}
D_{xy}^k \left(U,V\right) = \frac{U - \alpha_{xy} + (1-\tau_k) (V - \gamma_{xy} + w_k)}{2-\tau_k}.
\end{equation*}
\end{ex}

\subsection{Matching equilibrium}
\label{sec:matching_functions}

Now back to the matching model, recall that we had
\begin{equation*}
\frac{\mu_{xy}}{n_x} = \exp \left( \frac{U_{xy} - u_x}{\sigma} \right),
\qquad
\frac{\mu_{xy}}{m_y} = \exp \left( \frac{V_{xy} - v_y}{\sigma} \right),
\end{equation*}
where $U_{xy}$ and $V_{xy}$ denote the systematic parts of utility in a match $xy$.
Solving for $U_{xy}$ and $V_{xy}$ yields
\begin{equation*}
U_{xy} = u_x + \sigma \log \frac{\mu_{xy}}{n_x},
\qquad
V_{xy} = v_y + \sigma \log \frac{\mu_{xy}}{m_y}.
\end{equation*}
We assume that the market is large, so that every type of match $xy$ must be occurring.
Then $D_{xy} \left( U_{xy}, V_{xy} \right) = 0$ for all $xy$.
Replacing $U_{xy}$ and $V_{xy}$ by their expressions above, we get
\begin{equation*}
D_{xy} \left( u_x + \sigma \log \frac{\mu_{xy}}{n_x}, v_y + \sigma \log \frac{\mu_{xy}}{m_y} \right) = 0.
\end{equation*}
The translation invariance property yields 
\begin{equation*}
\sigma \log \mu_{xy} + D_{xy} \left( u_x - \sigma \log n_x, v_y - \sigma \log m_y \right) = 0.
\end{equation*}
Thus we have eliminated $w_{xy}$ by combining equations (\ref{eq:prob_x_picks_y}) and (\ref{eq:prob_y_picks_x}), and we are able to express $\mu_{xy}$, $\mu_{x0}$ and $\mu_{0y}$ as functions of $u_x$ and $v_y$ only:
\begin{equation*}
\mu_{xy} = \exp \left( -\frac{D_{xy} \left( u_x - \sigma \log n_x, v_y - \sigma \log m_y \right)}{\sigma} \right),
\end{equation*}
\begin{equation*}
\mu_{x0} = \exp \left( -\frac{u_x - \sigma \log n_x}{\sigma} \right),
\qquad
\mu_{0y} = \exp \left( -\frac{v_y - \sigma \log m_y}{\sigma} \right).
\end{equation*}

\bigskip

Next we want to solve for $u_x$ and $v_y$ using Jacobi's algorithm.
Introduce $p_x = -(u_x - \sigma \log n_x)$ and $p_y = v_y - \sigma \log m_y$, and substitute the expressions of $\mu_{xy}$, $\mu_{x0}$ and $\mu_{0y}$ above in the feasibility conditions (\ref{eq:feasibility_x_and_y}) to obtain the system:
\begin{equation}
\left\{ \;
\begin{aligned}[c]
&\textstyle\sum_y \exp \left( -D_{xy} \left( -p_x, p_y \right) /\sigma \right)
+ \exp \left( p_x / \sigma \right) = n_x \\
&\textstyle\sum_x \exp \left( -D_{xy} \left( -p_x, p_y \right) /\sigma \right)
+ \exp \left( -p_y / \sigma \right) = m_y.
\end{aligned}
\right.
\label{eq:matching_with_transfers_Jacobi_system}
\end{equation}
We define the function $Q$ using
\begin{equation*}
\left\{ \;
\begin{aligned}
Q_x \left(p\right)
&= \textstyle\sum_y \exp \left( - D_{xy} \left(-p_x, p_y \right) / \sigma \right)
+ \exp \left( p_x / \sigma \right) - n_x \\
Q_y \left(p\right)
&= -\textstyle\sum_x \exp \left( - D_{xy} \left(-p_x, p_y \right) / \sigma \right)
- \exp \left( -p_y / \sigma \right) + m_y,
\end{aligned}
\right.
\end{equation*}
so that the system (\ref{eq:matching_with_transfers_Jacobi_system}) rewrites as $Q \left(p\right) = 0$.
One can then check that $Q$ is a continuous M-function (it is a Z-function with strictly isotone aggregates).

We also have:

\begin{prop}
\label{prop:existence_sub_supersolution}
    The problem $Q \left(p\right) = 0$ has a sub- and a supersolution.
\end{prop}

\begin{proof}
To get a supersolution, %we will use the scarcity assumption.
first set $p_x$ such that $\exp \left( p_x / \sigma \right) \geq n_x$, so that $Q_x \left(p_x, p_{-x}\right) \geq 0$ for all $p_{-x}$.
Second, we show that for all $xy$ we must have $D_{xy}(-p_x, p_y) \to +\infty$ when $p_y \to +\infty$.
Suppose this is not the case: then there exists an increasing sequence $(p_y^n)$ such that $p_y^n \to +\infty$ but $D_{xy}(-p_x, p_y^n) \leq K$ for some constant $K$.
By translation invariance, $D_{xy}(-p_x-K, p_y^n-K) \leq 0$, i.e.\ $(-p_x-K, p_y^n-K) \in \mathcal F_{xy}$ for all $n$.
But since $-p_x-K$ is constant (so bounded below) and $p_y^n-K \to +\infty$, this contradicts the scarcity assumption.
Hence for $p_y$ large enough, $\textstyle\sum_x \exp \left( - D_{xy} \left(-p_x, p_y \right) / \sigma \right) + \exp \left( -p_y / \sigma \right) \leq m_y$, i.e.\ $Q_y \left(p\right) \geq 0$.
A subsolution can be found in a similar manner.
\end{proof}

Hence by theorem \ref{thm:Ortega-Rheinboldt1970}, the problem has a unique solution.
Finally, since the problem has a solution for any $n_x$ and $m_y$ the function $Q$ must be surjective, and therefore any Jacobi sequence converges towards the solution according to theorem \ref{thm:Ortega-Rheinboldt1970surjective}.

\bigskip

We conclude this discussion with a comment on the optimization structure.
If $Q$ were a gradient, one could write $Q_z = \partial F / \partial p_z$
for some function $F$ and thus
\begin{equation*}
\frac{\partial Q_y}{\partial p_x}
= \frac{\partial^2 F}{\partial p_x \partial p_y}
= \frac{\partial Q_x}{\partial p_y}.
\end{equation*}
But we have 
\begin{align*}
\frac{\partial Q_x \left(p\right)}{\partial p_y}
&= -\exp \left( -\frac{D_{xy} \left( -p_x, p_y \right) }{\sigma} \right)
\frac{\partial_V D_{xy} \left( -p_x, p_y \right)}{\sigma} \\
\frac{\partial Q_y \left(p\right) }{\partial p_x}
&= -\exp \left( -\frac{D_{xy} \left( -p_x, p_y \right) }{\sigma} \right)
\frac{\partial_U D_{xy} \left( -p_x, p_y \right)}{\sigma}
\end{align*}
which is not symmetric unless 
\begin{equation*}
\partial_U D_{xy} \left( U, V \right)
= \partial_V D_{xy} \left( U, V \right).
\end{equation*}
This only happens in the optimal transport case, when $D_{xy} \left(U,V\right) = (U + V - \Phi_{xy})/2$.

\subsection{Full assignment case}

Here we assume that $\sum_x n_x = \sum_y m_y$.
In the previous setting, define the map
$Q : \mathbb{R}^{X\cup Y\backslash \{ y_0\}} \rightarrow \mathbb{R}^{X\cup Y\backslash \{y_0\}}$ by
\begin{equation*}
\left\{ \;
\begin{aligned}
Q_x \left(p\right)
&= \textstyle\sum_y \exp \left( -D_{xy}(-p_x, p_y) \right) - n_x, \qquad x \in X \\
Q_y \left(p\right)
&= -\textstyle\sum_x \exp \left( -D_{xy}(-p_x, p_y) \right) + m_y, \quad y \in Y \backslash \{y_0\}
\end{aligned}
\right.
\end{equation*}
where we have normalized $p_{y_0} = \pi$.
To ensure that $Q$ is responsive in this setting (and therefore that Jacobi updates are well-defined) we will make an additional assumption on the feasible sets $\mathcal F_{xy}$:

\begin{ass}[Strong transferability]
For all $U$ there exists $V$ such that $(U,V) \in \mathcal{F}_{xy}$; and conversely, for all $V$ there exists $U$ such that $(U,V) \in \mathcal{F}_{xy}$.
\end{ass}

Under this assumption, any arbitrarily high level of utility is attainable for a partner within a match, as long as the other partner is willing to give up enough utility.

\citet*{ChooGalichonLiangWeber} show:

\begin{thm}
\label{thm:ChooGalichonLiangWeber}
The equation $Q \left(p\right) = 0$ has a solution.
\end{thm}

\begin{proof}
The proof is constructive and is done in three steps.
First, we look for a supersolution to initialize Jacobi's algorithm.
Second, we show that Jacobi updates are well-defined.
Finally, we show that the Jacobi sequence cannot diverge.

\underline{Step 1}.
Look for a supersolution $p^0$.
We have
\begin{equation*}
Q_x \left(p\right)
= \textstyle\sum_y \exp \left( -D_{xy}(-p_x, p_y) \right) - n_x
\geq \exp \left( -D_{xy_0}(-p_x, \pi) \right) - n_x.
\end{equation*}
By strong transferability, there exists $U_x$ such that $(U_x, \pi + \log n_x) \in \mathcal F_{xy}$, i.e.\ $D_{xy} (U_x, \pi + \log n_x) \leq 0$.
Taking $p_x^0 = \log n_x - U_x$, we have
$0 \geq D_{xy} (-p_x^0 + \log n_x, \pi + \log n_x) = D_{xy} (-p_x^0, \pi) + \log n_x$, i.e.\ $\exp \left(-D_{xy} (-p_x^0, \pi)\right) \geq n_x$, and therefore $Q_x \left(p_x^0, p_{-x}\right) \geq 0$ for all $p_{-x}$.
Next, by scarcity we have $D_{xy}(-p_x^0, p_y) \to +\infty$ as $p_y \to +\infty$ (see the argument in the proof of proposition \ref{prop:existence_sub_supersolution}) so we can set $p_y^0$ large enough such that
\begin{equation*}
\textstyle\sum_x \exp \left( -D_{xy}(-p_x^0, p_y^0) \right) \leq m_y,
\end{equation*}
hence $Q_y \left(p^0\right) \geq 0$ as well, so we have a supersolution.

\underline{Step 2}.
We now show that the Jacobi update from a supersolution is well-defined.
The function $p_x \mapsto \textstyle\sum_y \exp \left( -D_{xy}(-p_x, p_y^0) \right) - n_x$ is continuous, nonnegative for $p_x = p_x^0$, and by scarcity it converges to $-n_x < 0$ when $p_x \to -\infty$,
hence $p_x^1$ is well-defined.
%(since $D_{xy} (U,V) \to +\infty$ as $U \to +\infty$).

Now consider the function $p_y \mapsto -\textstyle\sum_x \exp \left( -D_{xy}(-p_x^0, p_y) \right) + m_y$.
It is continuous, and nonnegative for $p_y = p_y^0$.
Choose an $x$ arbitrarily, then by strong transferability, there exists $V_y$ such that $(-p_x^0 + \log m_y, V_y) \in \mathcal F_{xy}$. We have $0 \geq D_{xy} (-p_x^0 + \log m_y, V_y) = D_{xy} (-p_x^0, V_y - \log m_y) + \log m_y$, thus $\exp \left( -D_{xy} (-p_x^0, V_y - \log m_y) \right) \geq m_y$, and therefore the function above is negative for $p_y =  V_y - \log m_y$, hence $p_y^1$ is also well-defined.

By induction, $p^{t+1}$ is well-defined for all $t$ since $p^t$ is also a supersolution.

\underline{Step 3}.
Let us show that the Jacobi sequence initialized at the supersolution $p^0$ cannot diverge.
$Q$ is a Z-function with isotone aggregates, therefore it is an M\textsubscript{0}-function, and thus the Jacobi sequence $(p^t)$ starting from $p^0$ is a decreasing sequence of supersolutions.
Hence, either it converges to a solution, or it is unbounded.

However, because $p^t$ is a supersolution, we have 
\begin{equation*}
\textstyle\sum_x \exp \left( -D_{xy_0}(-p_x^t, \pi) \right) - m_{y_0}
= \textstyle\sum_{z \in X \cup Y \backslash \{y_0\}} Q_z \left(p^t\right) \geq 0
\end{equation*}
thus all the $p_x^t$ cannot go to $-\infty$.
Letting $x^*$ be such that 
$p_{x^*}^t \rightarrow p_{x^*}^\infty > -\infty$, we have 
\begin{equation*}
\exp \left( -D_{x^* y}(-p_{x^*}^t, p_y^t) \right)
\leq \textstyle\sum_x \exp \left( -D_{xy}(-p_x^t, p_y^t) \right) \leq m_y
\end{equation*}
thus all $p_y^t$ remain bounded below as well.
Finally, we have 
\begin{equation*}
\textstyle\sum_y \exp \left( -D_{xy}(-p_x^t, p_y^t) \right) \geq n_x
\end{equation*}
and thus all the $p_x^t$ remain bounded below too.
Hence the sequence converges.
\end{proof}

Uniqueness of the solution is not guaranteed under the strong transferability assumption, because then $Q$ is an M\textsubscript{0}-function.
To obtain uniqueness, one may ask an additional requirement on the feasible sets, which ensures that $Q$ is an M-function and therefore that theorem~\ref{thm:Ortega-Rheinboldt1970} applies.
For instance:

\begin{ass}[Strong local transferability]
For any $(U,V) \in \mathcal{F}_{xy}$ and any $\varepsilon > 0$, the open ball of center $(U,V)$ and of radius $\varepsilon$ contains both a point $(U',V')$ such that $U' > U$, and a point $(U'',V'') \in \mathcal{F}_{xy}$ such that $V'' > V$.
\end{ass}

This assumption rules out horizontal or vertical slopes on the frontier of $\mathcal F_{xy}$, which ensures that $Q$ is an M-function.

\subsection*{References and notes}
This section draws upon \cite*{galichon2019costly}.
The case with full assignment is covered in \citet*{ChooGalichonLiangWeber}.

\newpage

%\section{Lecture 3: models of matching without transfers}
\section{Models of matching without transfers}
\label{sec:notransfer}

In the previous section, utility could be (imperfectly) transferred within matches by adjusting the wage that the firm was paying to the worker.
In this section, we consider instead models in which each match leads to a fixed amount of utility for each partner.

\subsection{Introduction}

We consider a labor market with fixed wages.
Assume that $\mathcal{I}$ is the set of (individual) workers, and $\mathcal{J}$ is the set of (individual) firms.
A worker $i$ matched with firm $j$ gets utility $\alpha_{ij}$, and similarly, a firm $j$ matched with worker $i$ gets utility $\gamma_{ij}$.
Unassigned agents (workers or firms) get utility $0$.

We make an important assumption:

\begin{ass}[No indifference]
$\alpha_{ij} \neq 0$ and $\alpha_{ij} \neq \alpha_{ij'}$ for $j \neq j'$, and similarly, $\gamma_{ij} \neq 0$ and $\gamma_{ij} \neq \gamma_{i'j}$ for $i \neq i'$.
\end{ass}

\bigskip

We are looking at a model of one-to-one matching.
A matching $\mu = (\mu_{ij})$ is defined such that $\mu_{ij} \in \{0,1\}$ is equal to 1 if $i$ and $j$ are matched, and 0 otherwise.
$\mu_{i0}$ or $\mu_{0j}$ are respectively equal to $1$ if $i$ or $j$ are unassigned.
The condition for $\mu$ to be feasible is therefore
\begin{equation}
\left\{ \;
\begin{aligned}
\textstyle \sum_j \mu_{ij} + \mu_{i0} &= 1 \\ 
\textstyle \sum_i \mu_{ij} + \mu_{0j} &= 1.
\end{aligned}
\right.
\end{equation}

\begin{ex}
\label{ex:two_workers_one_firm}
Consider two workers and one firm.
Agents have a utility 1 if they are matched, and a utility 0 if they are not matched.
We see that both workers want to be matched with the firm;
however, the firm can only match with one worker.
Unlike the models with transfers from section \ref{sec:transfer}, there is no wage to clear the market.
As a result, at least one worker will not get their first choice.
\end{ex}

As seen in the previous example, we can follow two alternative approaches.
The first one requires to give up on the idea of equilibrium, which implies that individuals get their first choice.
In this case, we need a central planner to solve the market using an algorithm: this is Gale and Shapley's notion of a \emph{stable matching}, which can be computed by the \emph{deferred acceptance algorithm}
(see \citealt{gale1962college} and \citealt{roth1990two}).

The second method instead introduces a numéraire that cannot be transferred, like waiting times, to clear the market:
this is the approach introduced in \cite*{galichon2022monotone}.
This approach provides \emph{equilibrium matchings} that can be computed by the \emph{deferred acceptance with Lagrange multipliers (DALM) algorithm}.
%In the second method, instead, the markets are cleared by introducing a numéraire that cannot be transferred, like waiting times, which will be determined at equilibrium:

\bigskip

We will review both approaches in that order, and then study their relationships. 
In doing so we will also cover Adachi's reformulation of Gale and Shapley's stable matching problem, which allows us to use Gauss–Seidel's algorithm (closely related to Jacobi's algorithm) to solve the problem.

\subsection{Stable matchings}

Introduce $u_i$ and $v_j$ the respective payoffs obtained by worker $i \in \mathcal{I}$ and firm $j \in \mathcal{J}$ in the outcome of the game.

Note that if there were a pair $i$ and $j$ for which $u_i < \alpha_{ij}$ and $v_j < \gamma_{ij}$, then by matching together $i$ and $j$ could both obtain more utility than what is guaranteed to them in the outcome of the game.
This outcome would then not be stable, and $i$ and $j$ would form a \emph{blocking pair}.
Similarly, if $u_i < 0$, then $i$ would be better off remaining unmatched and getting utility 0 rather than taking the outcome payoff $u_i$.
A similar result holds on the other side of the market, and as a result we shall require that for all $i$ and $j$,
\begin{align*}
&\max \left( u_i - \alpha_{ij} , v_j - \gamma_{ij} \right) \geq 0, \\
&u_i \geq 0, \; v_j \geq 0,
\end{align*}
which are called the \emph{stability conditions}.

Finally, if $i$ and $j$ actually match, we expect $u_i$ to be equal to $\alpha_{ij}$ and $v_j$ to be equal to $\gamma_{ij}$.
Similarly, if $i$ or $j$ respectively remains unmatched, then $u_i$ or $v_j$ should be 0.
We call this a \emph{strong complementarity condition}, for reasons that will become clear later on.

\bigskip

To summarize, an outcome $(\mu, u, v)$ is \emph{stable} if: \begin{enumerate}

\item[(i)] $\mu$ is a feasible matching: $\mu_{ij} \in \{0,1\}$ and
\begin{align}
\textstyle \sum_j \mu_{ij} + \mu_{i0} &= 1 \label{eq:feasibility_i} \\ 
\textstyle \sum_i \mu_{ij} + \mu_{0j} &= 1, \label{eq:feasibility_j}
\end{align}

\item[(ii)] the stability conditions hold, that is
\begin{align}
&\max \left( u_i - \alpha_{ij} , v_j - \gamma_{ij} \right) \geq 0, \label{eq:stability_ij} \\
&u_i \geq 0, \; v_j \geq 0, \label{eq:stability_outside}
\end{align}

\item[(iii)] strong complementarity holds, that is%
\begin{align}
\mu_{ij} > 0 &\implies u_i = \alpha_{ij}, \; v_j = \gamma_{ij} \label{eq:complementarity_ij} \\
\mu_{i0} > 0 &\implies u_i = 0 \label{eq:complementarity_i} \\
\mu_{0j} > 0 &\implies v_j = 0. \label{eq:complementarity_j}
\end{align}

\end{enumerate}

\bigskip

Because of the strong complementarity conditions, one can deduce the utilities $u$ and $v$ of a stable outcome from its matching $\mu$.
This is in contrast with models of matching with transfers, in which different utility profiles can be compatible with the same matching.
For this reason, in models of matching without transfers we sometimes refer to \emph{stable matchings} instead of stable outcomes.

\bigskip

The standard method to find such a stable matching is Gale and Shapley's deferred acceptance algorithm.
The idea is as follows.
We keep track of a set of ``available'' offers that can be made by workers to firms;
this is initially unconstrained, which means that any worker can initially make an offer to any firm.
At each round, each worker makes an offer to their preferred firm among the set of those that are available to them.
If a firm receives several offers, it keeps its favorite and rejects the others.
Offers that have been rejected are no longer available.
The algorithm then iterates until no offer is rejected.

\bigskip

Formally, let's define $\mathcal A^t \left(i\right) \subseteq \mathcal J$ as the set of firms available to worker $i$ at time $t$,
$\mathcal P^t \left(i\right) \subseteq \mathcal J$ as the set of firms to whom worker $i$ makes an offer at time $t$ (either a singleton or empty),
and $\mathcal K^t \left(j\right) \subseteq \mathcal I$ as the set of offers kept by firm $j$ at the end of round $t$ (also either a singleton or empty).
Assume that at $t = 0$ all firms are available to anyone: $\mathcal{A}^0 \left(i\right) = \mathcal J$.

\bigskip

\noindent\rule{\textwidth}{0.4pt}
\textbf{Deferred acceptance algorithm} \citep{gale1962college}

Iterate over $t \geq 0$: \begin{enumerate}
\item Proposal phase:\vspace{-10pt}
\begin{equation*}
\mathcal P^t \left(i\right) = \arg\max_j \left\{ \alpha_{ij} : j \in \mathcal{A}^t \left(i\right) \right\}
\end{equation*}
or $\mathcal P^t \left(i\right) = \emptyset$ if the max is negative or $\mathcal{A}^t \left(i\right)$ is empty,
\vspace{5pt}
\item Disposal phase:\vspace{-10pt}
\begin{equation*}
\mathcal K^t \left(j\right) = \arg\max_i \left\{ \gamma_{ij} : j \in \mathcal P^t \left(i\right) \right\}
\end{equation*}
or $\mathcal K^t \left(j\right) = \emptyset$ if the max is negative or $j \notin \bigcup_i \mathcal P^t \left(i\right)$,
\vspace{5pt}
\item Adjustment phase:
\begin{equation*}
\mathcal{A}^{t+1} \left(i\right) = \mathcal{A}^t \left(i\right)
\backslash \left(\mathcal P^t \left(i\right) \backslash (\mathcal K^t)^{-1} \left(i\right) \right)
\end{equation*}
where $(\mathcal K^t)^{-1} \left(i\right)$ is the set of firms who kept worker $i$'s offer,

\end{enumerate}

until $\mathcal{A}^{t+1} \left(i\right) = \mathcal{A}^t \left(i\right)$.

\noindent\rule{\textwidth}{0.4pt}

\bigskip

\begin{thm}
The deferred acceptance algorithm converges to a stable matching.
\end{thm}

\begin{proof} The proof is deferred.
\end{proof}

\subsection{Adachi's formulation}

Adachi's formulation stems from a simple idea.
As argued above, in a stable matching individuals do not always get their first choice, since there is no numéraire to clear the market.
However, for each worker, one can define the \emph{consideration set} as the set of firms that are willing to match with that worker;
in a stable matching, each worker is matched with their preferred firm among the consideration set.
Similarly on the other side of the market, each firm is matched with their preferred worker among its consideration set.
Adachi's theorem states that this is actually a necessary and sufficient condition for a stable matching.

\bigskip

\begin{thm}[\citealt{adachi2000characterization}]
The outcome $(\mu,u,v)$ is stable if and only if
\begin{equation}
\left\{ \;
\begin{aligned}[c]
u_i &= \max_j \left\{ \alpha_{ij}, 0 : v_j \leq \gamma_{ij} \right\} \\
v_j &= \max_i \left\{ \gamma_{ij}, 0 : u_i \leq \alpha_{ij} \right\}.
\end{aligned}
\right.
\label{eq:Adachi}
\end{equation}
\end{thm}

Note that we use the shortened notation $\max_j \left\{ \alpha_{ij}, 0 : v_j \leq \gamma_{ij} \right\}$ to mean $\max(0, \max_j \left\{ \alpha_{ij} : v_j \leq \gamma_{ij} \right\} )$.

We can make some remarks.
First, this theorem yields a formulation of stable matchings as the fixed point of some operator -- we will come back to this below.
Second, the mapping $v \mapsto u$ is antitone with respect to $v$, and the mapping $u \mapsto v$ is antitone with respect to $u$.
Lastly, the function $u_i(v) = \max_j \left\{ \alpha_{ij} : v_j \leq \gamma_{ij} \right\}$ is reminiscent of the convex conjugate function $v_i^*(v) = \max_j \left\{\Phi_{ij} - v_j \right\}$.
However, $u = v^\ast$ and $v=u^\ast$ is not sufficient for $(u,v)$ to be solution to the dual optimal
transport problem.

\bigskip

\begin{proof}
For simplicity, we present the proof when $|\mathcal{I}| = |\mathcal{J}|$ and all $\alpha$ and $\gamma $ are positive, so that in a stable outcome everyone must be matched.
For the direct implication, assume that $(\mu,u,v)$ is stable.
Looking for a contradiction, assume that (\ref{eq:Adachi}) does not hold: without loss of generality,
\begin{equation*}
u_i \neq \max_j \left\{ \alpha_{ij} : v_j \leq \gamma_{ij} \right\}
\quad \text{for some $i$.}
\end{equation*}
Worker $i$ is matched with $J \left(i\right)$, so that $u_i = \alpha_{i J\left(i\right)}$ and $v_{J\left(i\right)}= \gamma_{i J\left(i\right)}$.
As a result
\begin{equation*}
\max_j \left\{ \alpha_{ij} : v_j \leq \gamma_{ij} \right\} \geq \alpha_{i J\left(i\right)} = u_i
\end{equation*}
thus we have $\max_j \left\{ \alpha_{ij} : v_j \leq \gamma_{ij} \right\} > u_i = \alpha_{i J\left(i\right)}$.
Denote $j^* \neq J \left(i\right)$ the corresponding maximizer,
we have $v_{j^*} \leq \gamma_{ij^*}$ and $\alpha_{ij^\ast} > u_i$.
But no indifference implies $v_{j^*} < \gamma_{ij^*}$, therefore $i$ and $j^*$ form a blocking pair.
Contradiction, thus (\ref{eq:Adachi}) in fact holds.

Conversely, assume that the equalities (\ref{eq:Adachi}) hold.
Define $\mu_{ij} = 1$ iff $u_i = \alpha_{ij}$.
Since $u_i = \alpha_{ij}$, $j$ must be in the feasible set for $i$, i.e.\ $v_j \leq \gamma_{ij}$.
Furthermore, $i$ is such that $u_i \leq \alpha_{ij}$, thus $v_j \geq \gamma_{ij}$.
As a result, $v_j = \gamma_{ij}$.
By no indifference, it also means that the matching $\mu$ thus defined is feasible.

Let us show that $(\mu,u,v)$ is Gale--Shapley stable.
Assume $ij$ is a blocking pair, so that $\alpha_{ij} > u_i$ and $\gamma_{ij} > v_j$.
But $\gamma_{ij} > v_j$ implies $v_j \leq \gamma_{ij}$, thus $u_i \geq \alpha_{ij}$, which is a contradiction.
\end{proof}

\bigskip

\citet{GalichonLegerZmappings} build on Adachi's formulation to rewrite the stability problem as a model with substitutes, specifically in the form $Q(p) = 0$ with $Q$ an M\textsubscript{0}-function.

Use the change-of-sign trick in (\ref{eq:Adachi}): set $p_i = -u_i$ and $p_j = v_j$, and define Adachi's map $T$ as
\begin{equation*}
\left\{ \;
\begin{aligned}[c]
T_i \left(p\right)
&= \min_j \left\{ -\alpha_{ij}, 0 : \gamma_{ij} \geq p_j \right\} \\
T_j \left(p\right)
&= \max_i \left\{ \gamma_{ij}, 0 : p_i \geq -\alpha_{ij} \right\} .
\end{aligned}
\right.
\end{equation*}

Clearly $T$ is an isotone map, and by Adachi's theorem we know that $(\mu,u,v)$ is stable if and only if $p = (-u,v)$ satisfies the fixed-point equation
\begin{equation*}
p = T \left(p\right).
\end{equation*}
Even though $p - T \left(p\right)$ is a Z-function, it is not an M\textsubscript{0}-function in general.
We set instead
\begin{equation*}
\left\{ \;
\begin{aligned}[c]
M_{ij} \left(p\right)
&= 1 \left\{ p_i \geq -\alpha_{ij}, \gamma_{ij} \geq p_j\right\} \\
M_{i0} \left(p\right)
&= 1 \left\{ p_i \geq 0 \right\} \\
M_{0j} \left(p\right)
&= 1 \left\{ 0 \geq p_j \right\}
\end{aligned}
\right.
\end{equation*}
and we define the function $Q$ with
\begin{equation*}
\left\{ \;
\begin{aligned}[c]
Q_i \left(p\right)
&= \textstyle\sum_{j \in \mathcal{J}} M_{ij} \left(p\right) + M_{i0} \left(p\right) - 1 \\
Q_j \left(p\right)
&= -\textstyle\sum_{i \in \mathcal{I}} M_{ij} \left(p\right) - M_{0j} \left(p\right) + 1.
\end{aligned}
\right.
\end{equation*}

\bigskip

% Q is not continuous but 'almost'? i.e. if we move p_i by epsilon we move Q_i by at most 1
Although $Q$ is not continuous, we have:

\begin{thm}[\citealt{GalichonLegerZmappings}] ~ \begin{enumerate}
\item[(i)] $Q$ is an M\textsubscript{0}-function.
\item[(ii)] Adachi's map $T$ is the coordinate update operator associated with $Q$.
\end{enumerate}
\end{thm}

\bigskip

As a result, $T \left(p\right) = p$ if and only if $Q \left(p\right) = 0$, and thus the set of zeroes of $Q$ is a sublattice of $\mathbb{R}^n$ (cf.\ corollary of theorem \ref{thm:inverse_isotonicity}).
Hence:

\begin{coro}[\citealt{Knuth1997}]
Given two stable outcomes $(\mu,u,v)$ and $(\mu',u',v')$, define
\begin{equation}
\left\{ \;
\begin{aligned}[c]
(\mu \wedge_{\mathcal{I}} \mu')_{ij}
&= 1 \left\{ u_i \leq u_i' \right\} \mu_{ij} + 1 \left\{ u_i > u_i' \right\} \mu_{ij}' \\
(\mu \vee_{\mathcal{I}} \mu')_{ij}
&= 1 \left\{ u_i > u_i' \right\} \mu_{ij} + 1 \left\{ u_i \leq u_i' \right\} \mu_{ij}'.
\end{aligned}
\right.
\end{equation}
Then $\mu \wedge_{\mathcal{I}} \mu'$ and $\mu \vee_{\mathcal{I}} \mu'$ are stable matchings, and
\begin{equation}
\left\{ \;
\begin{aligned}[c]
\mu \wedge_{\mathcal{I}} \mu'
&= \mu \vee_{\mathcal{J}} \mu' \\
\mu \vee_{\mathcal{I}} \mu'
&= \mu \wedge_{\mathcal{J}} \mu'.
\end{aligned}
\right.
\end{equation}
\end{coro}

\bigskip

As a result, the lattice of stable matchings has an element which is preferred by all the $i$'s and least liked by the $j$'s;
and an element which is preferred by all the $j$'s and least liked by the $i$'s.

In other words, workers $i$ and firms $j$ have \emph{opposite interests}: what is better for workers is worse for firms and conversely.

\bigskip

Adachi's algorithm is nothing else than Gauss–Seidel's algorithm applied to finding the zeroes of $Q$. (Gauss–Seidel's algorithm is closely related to Jacobi's algorithm, the difference being that coordinate updates are made sequentially with Gauss–Seidel, instead of in parallel with Jacobi.)

\bigskip

\noindent\rule{\textwidth}{0.4pt}
\textbf{Adachi's algorithm}

Set $p^0$ low enough, for instance $p_i^0 = \min_j \left\{ -\alpha_{ij}, 0 \right\}$ and $p_j^0 = \min_i \left\{ \gamma_{ij}, 0 \right\}$.
Iterate over $t \geq 0$:
\begin{equation}
\left\{ \;
\begin{aligned}[c]
p_i^{t+1} &= \min_j \left\{ -\alpha_{ij}, 0 : \gamma_{ij} \geq p_j^t \right\} \\
p_j^{t+1} &= \max_i \left\{ \gamma_{ij}, 0 : p_i^{t+1} \geq - \alpha_{ij} \right\}
\end{aligned}
\right.
\end{equation}
until $p^{t+1} = p^t$.

\noindent\rule{\textwidth}{0.4pt}

\bigskip

Note that if it starts from a subsolution, Adachi's algorithm will return the $\mathcal{I}$-preferred stable matching,
and if it starts from a supersolution, it will return the $\mathcal{J}$-preferred stable matching.

\bigskip

In Gale and Shapley's deferred acceptance algorithm, at each period $u_i^t$ can only decrease by one step in the scale of the rankings induced by $\alpha_{ij}$.
Indeed, if the offer made by worker $i$ is kept, then $i$ keeps making the same offer;
while if the offer made by $i$ is turned down, then $i$ will move to their next preferred firm.
This may however be inefficient, as $i$ is allowed to make an offer to a firm $j$ who already has a dominating offer, and who will thus turn down $i$.

A more efficient version of deferred acceptance consists in modifying Gale and Shapley's algorithm in order to induce the workers $i$ whose offer has been turned down to move to the next firm $j$ in their preference list which does not already have an offer better than $i$.
This is exactly what Adachi's algorithm does.

\bigskip

Formally, let us denote 
\begin{equation}
\left\{ \;
\begin{aligned}[c]
N_i \left(p\right)
&= \max_j \left\{ -\alpha_{ij} : -\alpha_{ij} < p_i \right\} \\
N_j \left(p\right)
&= \max_i \left\{ \gamma_{ij} : \gamma_{ij} < p_j \right\}
\end{aligned}
\right.
\end{equation}
which looks for the match just below the match implied by $p$ for the $i$'s,
and the match just above the match implied by $p$ for the $j$'s.

Gale and Shapley's algorithm can be interpreted as
\begin{equation}
\left\{ \;
\begin{aligned}[c]
p_i^{2t+1} &= \min \left\{ T_i \left(p^{2t}\right), N_i \left(p^{2t}\right) \right\} \\
p_j^{2t+2} &= T_j \left(p^{2t+1}\right)
\end{aligned}
\right.
\end{equation}
which is a ``damped Gauss--Seidel'' algorithm (where $T$ is Adachi's map).

\subsection{Equilibrium matchings}

The notion of equilibrium matchings was introduced by \citet*{galichon2023monotone}.
Just like stable matchings, equilibrium matchings require ruling out blocking pairs, so the stability conditions will be the same.
The only difference is that we shall not require that if $i$ and $j$ are matched, then $u_i = \alpha_{ij}$ and $v_j = \gamma_{ij}$, but instead that $u_i \leq \alpha_{ij}$ and $v_j \leq \gamma_{ij}$ with at most one of these inequalities being strict.
That is, we are fine with burning utility on one side of the market provided that we don't burn it on both sides.
Hence the strong complementarity condition (\ref{eq:complementarity_ij}) becomes a weak one, only requiring that
\begin{equation}
\mu_{ij} > 0 \implies \max \left( u_i - \alpha_{ij}, v_j - \gamma_{ij} \right) = 0.
\label{eq:weak_complementarity_ij}
\end{equation}

An outcome $(\mu,u,v)$ is thus an \emph{equilibrium matching} if it satisfies conditions (\ref{eq:feasibility_i})--(\ref{eq:complementarity_j}), but replacing the strong complementarity condition (\ref{eq:complementarity_ij}) with (\ref{eq:weak_complementarity_ij}).

\bigskip

Of course, if an outcome is stable, then it is an equilibrium matching.
We have seen that stable outcomes exist, so why is the notion of an equilibrium matching useful?
The reason is that we can define a notion of aggregate equilibrium matching, but not of aggregate stable matching.

\bigskip

We will now discard the no-indifference assumption by considering a population similar to the one we studied in the matching with transfers setting.
Workers and firms are sorted into types $x \in \mathcal X$ and $y \in \mathcal Y$ respectively, so that there are $n_x$ identical workers of type $x$, and $m_y$ identical firms of type $y$.
The value of the match $xy$ is $\alpha_{xy}$ for a worker $x$, and $\gamma_{xy}$ for a firm $y$.
Hence workers $x$ are indifferent between all firms $y$, and reciprocally.
Example \ref{ex:two_workers_one_firm} showed that in this case, aggregate stable matchings may not necessarily exist since two workers $x$ might end up with different utilities depending on which firm they are matched with, when we want to be able to define $u_x$ the ex-post utility for all workers $x$.
In contrast, aggregate equilibrium matchings do exist.

\begin{defn} An outcome $(\mu,u,v)$ is an \emph{aggregate equilibrium matching} if: \begin{enumerate}

\item[(i)] $\mu$ is a feasible matching: $\mu_{xy} \geq 0$, and
\begin{align}
\textstyle \sum_y \mu_{xy} + \mu_{x0} &=n_x \\ 
\textstyle \sum_x \mu_{xy} + \mu_{0y} &= m_y,
\end{align}

\item[(ii)] the stability conditions hold, that is
\begin{align}
&\max \left( u_x - \alpha_{xy}, v_y - \gamma_{xy} \right) \geq 0 \\
&u_x \geq 0, \; v_y \geq 0,
\end{align}

\item[(iii)] weak complementarity holds, that is
\begin{align}
\mu_{xy} > 0 &\implies \max \left( u_x - \alpha_{xy}, v_y - \gamma_{xy} \right) = 0
\label{eq:weak_complementarity_xy} \\
\mu_{x0} > 0 &\implies u_x = 0 \\
\mu_{0y} > 0 &\implies v_y = 0.
\end{align}
\end{enumerate}

\end{defn}

\bigskip

It is clear that this definition coincides with the notion of (individual) equilibrium matching.
We have just replaced the margins and allowed $\mu_{xy}$ to take other values than just 0 or 1.
This is part of the theory seen in section \ref{sec:transfer} with the distance function
\begin{equation}
D_{xy} \left( U,V \right) = \max \left( U - \alpha_{xy}, V - \gamma_{xy} \right).
\label{eq:distance_notransfer}
\end{equation}

\bigskip

To compute aggregate equilibrium matchings, \citet*{galichon2023monotone} consider an aggregate version of Gale and Shapley's deferred acceptance algorithm.
Let $\mu_{xy}^{A,t}$ denote the number of positions of type $y$ available to workers of type $x$ at time $t$.
At $t=0$ all positions are open: $\mu_{xy}^{A,0} = \min \left\{ n_x, m_y \right\}$.

\bigskip

\noindent\rule{\textwidth}{0.4pt}
\textbf{Deferred Acceptance with Lagrange multipliers (DALM)}

Define $\mu_{xy}^{A,0} = \min \left\{ n_x, m_y \right\}$, and iterate over $t \geq 0$: \begin{enumerate}

\item Proposal phase: \vspace{-10pt}
\begin{align*}
\mu^{P,t} \in \arg \max_\mu ~ & \sum_{xy} \mu_{xy} \alpha_{xy} \\
\text{s.t.}~ &\sum_y \mu_{xy} \leq n_x \\
&\mu_{xy} \leq \mu_{xy}^{A,t}
\end{align*}

\item Disposal phase: \vspace{-10pt}
\begin{align*}
\mu^{K,t} \in \arg\max_\mu ~ & \sum_{xy} \mu_{xy} \gamma_{xy} \\
\text{s.t.}~ &\sum_x \mu_{xy} \leq m_y \\
&\mu_{xy} \leq \mu_{xy}^{P,t}
\end{align*}

\item Adjustment phase: \vspace{-10pt}
\begin{equation*}
\mu_{xy}^{A,t+1} = \mu_{xy}^{A,t} - (\mu_{xy}^{P,t} - \mu_{xy}^{K,t})
\end{equation*}

\end{enumerate}

until $\mu_{xy}^{A,t+1} = \mu_{xy}^{A,t}$.

\noindent\rule{\textwidth}{0.4pt}

\bigskip

In the proposal and disposal phases, there are Lagrange multipliers $\tau_{xy}^\alpha \geq 0$ and $\tau_{xy}^\gamma \geq 0$ which are such that the perceived utilities are respectively $\alpha_{xy} - \tau_{xy}^\alpha$ for workers, and $\gamma_{xy} - \tau_{xy}^\gamma$ for firms.
We can then show that $\tau_{xy}^\alpha$ is increasing and $\tau_{xy}^\gamma$ is decreasing.

\bigskip

We have the following result:

\begin{thm}[\citealt*{galichon2023monotone}]
Aggregate equilibrium matchings exist, and DALM returns one of them.
\end{thm}

\bigskip

The proof relies on novel comparative statics for \emph{exchangeable functions} -- which are not covered by Topkis' and Veinott's monotone comparative statics theory.

\subsection{Application: housing market with rent control}

Consider a housing market in which the rent is capped.
Renters $i$ are categorized according to family size $x \in X$.
Houses $j$ are categorized according to the number of rooms $y \in Y$.
We assume that the rent for a house $y$ is capped at $r_y$ and that houses are under-supplied in the market, so that in practice $r_y$ is the going rent for any house $y$.

The utility of a family $i$ of size $x$ renting a house $y$ is $a_{xy} - r_y + \varepsilon_{iy}$,
where $a_{xy}$ represents the monetary value of the amenities of a house $y$ for a family of size $x$,
and $\varepsilon_{iy}$ is a random utility component.
If unmatched, the family gets utility $\varepsilon_{i0}$.
The utility of a landlord $j$ who rents out a house $y$ to a family of size $x$ is $r_y - c_{xy} + \eta_{xj}$,
where $c_{xy}$ is the maintenance cost which depends on the family size and house type,
and $\eta_{xj}$ is also random.
If unmatched, the landlord gets utility $\eta_{0j}$.

We can let $\alpha_{xy} = a_{xy} - r_y$ and $\gamma_{xy} = r_y - c_{xy}$, and the analysis follows the same steps as in section \ref{sec:transfer}.
The respective problems of family $i$ and landlord $j$ are
\begin{equation*}
\max_y \, \{\alpha_{xy} - t_{xy} + \varepsilon_{iy}, \varepsilon_{i0}\},
\qquad
\max_x \, \{\gamma_{xy} - s_{xy} + \eta_{xj}, \eta_{0j}\}
\end{equation*}
where $t_{xy}, s_{xy} \geq 0$ are the ``burnt utilities'' on each side of the match.

Assuming the random utility components are i.i.d.\ Gumbel, the expected indirect utilities for a family $x$ and a landlord $y$ are
\begin{equation*}
u_x = \log \bigg(1 + \sum_y \exp \left( \alpha_{xy} - t_{xy} \right)\bigg),
\qquad
v_y = \log \bigg(1 + \sum_x \exp \left( \gamma_{xy} - s_{xy} \right)\bigg)
\end{equation*}
and the probabilities that a family of type $x$ picks a house $y$, and that a landlord $y$ chooses a family $x$, are respectively
\begin{equation*}
\frac{\mu_{xy}}{n_x} = \exp \left( \alpha_{xy} - t_{xy} - u_x \right),
\qquad
\frac{\mu_{xy}}{m_y} = \exp \left( \gamma_{xy} - s_{xy} - v_y \right).
\end{equation*}

\bigskip

With the large market assumption, so that every type of match occurs in equilibrium, utility can only be burnt on one side of each match $xy$, so that
\begin{equation*}
\min(t_{xy}, s_{xy}) = 0.
\end{equation*}
Substituting $t_{xy}$ and $s_{xy}$ from the equations above, we obtain
\begin{equation*}
\log \mu_{xy}
= \min \big( \alpha_{xy} - u_x + \log n_x, \, \gamma_{xy} - v_y + \log m_y \big).
\end{equation*}
This corresponds exactly to what we had in section \ref{sec:transfer}, but adapted to the distance-to-frontier function in (\ref{eq:distance_notransfer}).
Using $\mu_{x0} = \exp(-u_x + \log n_x)$ and $\mu_{0y} = \exp(-v_y + \log m_y)$, this rewrites as
\begin{equation*}
\mu_{xy}
= \min \big\{ \mu_{x0} \exp \left( \alpha_{xy} \right), \mu_{0y} \exp \left( \gamma_{xy} \right) \big\}.
\end{equation*}
Hence $\mu_{x0}$ and $\mu_{0y}$ must satisfy the system
\begin{equation*}
\left\{ \;
\begin{aligned}
\mu_{x0} + \sum_y \min \big\{ \mu_{x0} \exp \left( \alpha_{xy} \right), \mu_{0y} \exp \left( \gamma_{xy} \right) \big\} &= n_x
\\
\mu_{0y}+\sum_x \min \big\{ \mu_{x0}\exp \left( \alpha_{xy} \right), \mu_{0y} \exp \left( \gamma_{xy} \right) \big\} &= m_y.
\end{aligned}
\right.
\end{equation*}

Defining $p$ with $\mu_{x0} = e^{p_x}$ and $\mu_{0y} = e^{-p_y}$, and $Q$ with
\begin{equation*}
\left\{ \;
\begin{aligned}
Q_x \left(p\right)
&= e^{p_x} + \textstyle\sum_y \min \big\{ e^{p_x + \alpha_{xy}}, e^{\gamma_{xy} - p_y} \big\} - n_x \\
Q_y \left(p\right)
&= - e^{-p_y} - \textstyle\sum_x \min \big\{ e^{p_x + \alpha_{xy}}, e^{\gamma_{xy} - p_y} \big\} + m_y,
\end{aligned}
\right.
\end{equation*}
the system rewrites as $Q \left(p\right) = 0$, and one can check that $Q$ is a continuous Z-function with strictly isotone aggregates (hence an M-function).
The general theory applies: there exist both a sub- and a supersolution (obtained respectively by setting all prices very low or very high), and therefore the equilibrium exists and is unique.

In the full assignment case with $\sum_x n_x = \sum_y m_y$ and no outside option, we normalize $p_{y_0} = \pi$ and the problem becomes
\begin{equation*}
\left\{ \;
\begin{aligned}
Q_x \left(p\right)
&= \textstyle\sum_y \min \big\{ e^{p_x + \alpha_{xy}}, e^{\gamma_{xy} - p_y} \big\} - n_x,
\quad x \in X\\
Q_y \left(p\right)
&= - \textstyle\sum_x \min \big\{ e^{p_x + \alpha_{xy}}, e^{\gamma_{xy} - p_y} \big\} + m_y,
\quad y \in Y \backslash \{y_0\}.
\end{aligned}
\right.
\end{equation*}
Here $Q : \mathbb{R}^{X \cup Y \backslash \{y_0\}} \to \mathbb{R}^{X \cup Y \backslash \{y_0\}}$ is a Z-function with isotone aggregates (hence an M\textsubscript{0}-function).
However, because of the $\min$'s, $Q$ may not be responsive for some values of $p$, and for that reason it is unclear whether the Jacobi sequence is well-defined or not in this case.
Finding conditions under which the Jacobi algorithm works in the full assignment case with non-transferable utility thus remains an open problem for now.

\subsection{References and notes}

This section draws upon the book by \cite{roth1990two}.
It also builds on the papers by \citet*{galichon2019costly}, \cite{adachi2000characterization} and \cite{gale1962college}.
Additional references include the working papers by \citet*{GalichonLegerZmappings}, and \citet*{galichon2023monotone}.

\clearpage

\bibliographystyle{agsm}
\bibliography{refs.bib}

\end{document}